\documentclass[a4paper,12pt]{article}
\pdfoutput=1

\usepackage{float}

\usepackage{amssymb} 
\usepackage{amsmath}
\usepackage{epsfig}
\usepackage{graphicx}
\usepackage{tikz}

\makeatletter

\@addtoreset{equation}{section}
\makeatother
\def\be{\begin{equation}}
\def\ee{\end{equation}}
\def\bea{\begin{eqnarray}}
\def\eea{\end{eqnarray}}

\def\({\left(}
\def\){\right)}
\def\<{\left<}
\def\>{\right>}

\def\be{\begin{equation}}
\def\ee{\end{equation}}

\def\ben{\begin{eqnarray}}
\def\een{\end{eqnarray}}
\def\({\left(}
\def\){\right)}
\def\<{\left<}
\def\>{\right>}
\def\!{\right|}
\def\|{\left|}

\def\[{\left[}
\def\]{\right]}

\def\+{\bar}

\def\O{{\cal{O}}}

\def\eps{{\cal{\varepsilon}}}

\begin{document}

\setlength{\unitlength}{1mm}

\pagestyle{empty}
\vskip-10pt
\vskip-10pt
\hfill 
\begin{center}
\vskip 3truecm
{\Large \bf
Wilson surface correlator 
\vskip 0.6truecm
in AdS/CFT
}
\vskip 2truecm
{\large \bf
Andreas Gustavsson}
\vspace{1cm} 
\begin{center} 
Physics Department, University of Seoul, Seoul 02504 KOREA
\end{center}
\end{center}
\vskip 1truecm
\vskip 0.5truecm
{\abstract We use the AdS/CFT correspondence to show that for two spherical Wilson surfaces, both in the fundamental representation of $SU(N)$ gauge group, in the six-dimensional $(2,0)$ theory, in the large $N$ limit there is a first order Gross-Ooguri phase transition at a certain critical separation distance that we obtain numerically.}

\vfill
\vskip4pt
\eject
\pagestyle{plain}

\section{Introduction}
That there will be a first order phase transition in the correlation function of two concentric and separated circular Wilson loops in four-dimensional maximally supersymmetic Yang-Mills theory as we increase the separation between the two Wilson loops beyond a certain critical value was first conjectured by Gross and Ooguri in \cite{Gross:1998gk}. But they used an imprecise argument of an Euler catenoid in flat space instead of a minimal surface in AdS space. The Gross-Ooguri phase transition was shortly thereafter demonstrated by Zarembo \cite{Zarembo:1999bu} using a partially analytic and partially numerical approach to analyze the minimal surface in AdS space. This result was further extended in \cite{Olesen:2000ji} and in \cite{Kim:2001td} an exact analytic relation was discovered. 

A natural next step would be to generalize to two spherical Wilson surfaces in the six-dimensional $(2,0)$ theory and examine whether there is a Gross-Ooguri phase transition at a certain critical separation distance. In AdS space the generalization is straightforward although we know little about the fundamentals of this mysterious six-dimensional theory. In \cite{Liu:2013uoa} a possible Gross-Ooguri transition was studied for the case of one spherical Wilson surface in the fundamental representation and the other spherical Wilson in a high-rank representation of the $SU(N)$ gauge group. This system was studied in AdS space with a five-brane in the bulk. In this paper we will study the more basic problem where both Wilson surfaces are in the fundamental representation. This system is described on the AdS side by a membrane whose volume we shall minimize. 
  
The correlation function between two separated Wilson surfaces can be presented explicitly for an abelian gauge group where we have an explicit formula for the Wilson surface expectation value. If we are given a closed surface $\Sigma$ then the abelian Wilson surface 
\bea
W(\Sigma) &=& e^{i e \int_{\Sigma} B}
\eea
gets the expectation value 
\bea
\<W(\Sigma)\> &=& e^{- e^2 \int_{x\in \Sigma} \int_{x'\in \Sigma} \<B(x) B(x')\>}
\eea
For the Wilson surface in the $(2,0)$ theory we also need to extract the selfdual contribution, but here we will not do this. Let us consider a surface that is a disconnected union of two surfaces,
\bea
\Sigma &=& \Sigma_1 \cup \Sigma_2
\eea
In that case, the expectation value can be expanded as
\bea
\<W(\Sigma)\> &=& \<W(\Sigma_1)\> \<W(\Sigma_2)\> \<W(\Sigma_1) W(\Sigma_2)\>
\eea
where we define
\bea
\<W(\Sigma_1) W(\Sigma_2)\> &:=& e^{- e^2 \int_{x\in \Sigma_1} \int_{x'\in \Sigma_2} \<B(x) B(x')\>} 
\eea
as the correlation function between the two Wilson surfaces. From the above, we see that the correlation function is given by the ratio
\bea
\<W(\Sigma_1)W(\Sigma_2)\> &=&  \frac{\<W(\Sigma)\>}{\<W(\Sigma_1)\> \<W(\Sigma_2)\>}
\eea
In the AdS/CFT correspondence, for $SU(N)$ gauge group at large $N$, we can compute the expectation value of a Wilson surface by computing the volume $V(\Sigma)$ of a minimal hypersurface in AdS that ends on the surface $\Sigma$ on the boundary,
\bea
\<W(\Sigma)\> &=& e^{-\frac{2N}{\pi} V(\Sigma)}
\eea
More precisely, the volume is infinite and needs to be regularized. We do this the simplest possible way by letting the surface end a short coordinate distance $z = \eps$ away from the boundary where $z$ denotes the Poincare coordinate that goes into the bulk. In AdS space we compute the correlation function as
\bea
\<W(\Sigma_1) W(\Sigma_2)\> &=& e^{- \frac{2N}{\pi}\(V(\Sigma) - V(\Sigma_1) - V(\Sigma_2)\)}
\eea
Here $V(\Sigma)$ denotes the volume of the connected minimal hypersurface whose boundary is $\Sigma = \Sigma_1 \cup \Sigma_2$, $V(\Sigma_1)$ denotes the volume of the minimal hypersurface whose boundary is $\Sigma_1$ and $V(\Sigma_2)$ denotes the volume of the minimal hypersurface whose boundary is $\Sigma_2$. Thus we shall compute a difference of volumes,
\bea
V(\Sigma) - V(\Sigma_1) - V(\Sigma_2)
\eea
This volume difference is a physical quantity. By taking the difference, both the $1/\eps^2$ and log($\eps$) divergences cancel and we are left with a finite physical volume difference. In this volume difference there is also no conformal anomaly. We can argue for this by noting that log($\eps$) terms cancel and by knowing that the conformal anomaly is closely tied to this log divergence. Another way to argue for the absence of a conformal anomaly is by returning to the abelian correlation function and recalling how we compute the correlation between two different surfaces by evaluating a propagator from one point of the first surface to another point on the other surface. The separation distance between these points is huge, compared to any infinitesimal separation of two points, simply by the fact that the two surfaces are separated from each other. The conformal anomaly, on the other hand, is computed locally by evaluating the propagator for two infinitesimally nearby points. Simply because we never extract such a short-distance propagator, it is clear that we can not get a conformal anomaly when we compute a correlator between two distinct surfaces. 

We will write the volume of the connected surface as
\bea
V_E(\Sigma) := V(\Sigma)
\eea
with a subscript $E$ and the volume of the disconnected surfaces as
\bea
V_G(\Sigma) := V(\Sigma_1) + V(\Sigma_2)
\eea
with a subscript $G$. We will use these subscripts $E$ and $G$ that refer to Euler and Goldschmidt respectively. In 1744 Euler presented the connected catenoid solution between two circles in flat space. Later Goldschmidt showed that there will be also disconnected disk solutions inside each of these two circles. Both the connected Euler catenoid and the disconnected Goldschmidt disk solutions extremize the volume. For a short separation distance between the two circles, the Euler catenoid has a smaller area, $A_E < A_G$ while for a sufficiently large separation distance the Goldschmidt solution has a smaller area, $A_G<A_E$. At a critical distance the two solutions have the same area, $A_E = A_G$. 

Let us now return to the realization of the Wilson surface as a minimal hypersurface in AdS space. For a spherical Wilson surface the minimal hypersurface solution and its volume were obtained by Berenstein, Corrado, Fischler and Maldacena in \cite{Berenstein:1998ij} where they used a trick whereby they conformally mapped the plane into the sphere and found an extension of this map to the entire AdS bulk space. The expectation value for an abelian spherical Wilson surface can be also computed and this computation was first done by E. Flink in his master's thesis. The result can be found in for instance \cite{Gustavsson:2004gj}. 

In the AdS dual picture that is valid at large $N$ there are in the AdS space two competing membrane saddles. One saddle is what we will refer to as the AdS-Goldschmidt saddle. This is the saddle point solution that consists of two copies of the minimal hypersurface corresponding to a single spherical Wilson surface of \cite{Berenstein:1998ij}. The other saddle is what we will refer to as the AdS-Euler saddle. This is the hypersurface solution that we will obtain partially analytically and partially numerically in this paper. 

In a more precise argument, the expectation value of two spherical Wilson surfaces in the large $N$ limit is a sum over all the saddle points -- all extremal hypersurfaces in AdS whose boundaries are at the Wilson surface,
\bea
\<W(\Sigma)\> &=& D_G e^{- \frac{2 N}{\pi} V_G} + D_{E_{stable}} e^{- \frac{2 N}{\pi} V_{E_{stable}}} + D_{E_{unstable}} e^{- \frac{2 N}{\pi} V_{E_{unstable}}}
\eea
where $D_G$, $D_{E_{stable}}$ and $D_{E_{unstable}}$ represent contributions from the quantum fluctuations around these saddles -- in the Gaussian approximation these are the one-loop determinants. Here $V_G$ denotes the volume of the AdS-Goldschmidt solution, $V_{E_{stable}}$ denotes the volume of the stable AdS-Euler solution and $V_{E_{unstable}}$ denotes the volume of the unstable AdS-Euler solution. Following \cite{Berenstein:1998ij}, we use units where the radius of the AdS space is one. There are actually two kinds of AdS-Euler surfaces both of which are solving the equations of motion. One solution is stable and the other is unstable. We will not consider the unstable AdS-Euler saddle any further and will henceforth use the short-hand notation $V_E := V_{E_{stable}}$ for its volume. For a small separation the stable AdS-Euler solution has a smaller volume than the AdS-Goldschmidt solution,
\bea
V_E < V_G
\eea
We may then write the partition function as
\bea
Z &=& D_E e^{-\frac{2N}{\pi} V_E} \(1 + \frac{D_G}{D_E} e^{- \frac{2N}{\pi} \(V_G-V_E\)}\)
\eea
if we neglect the contribution of the unstable AdS-Euler solution. The dominating contribution is that of the stable AdS-Euler solution. Subleading terms are exponentially suppressed and hence completely negligible in the large $N$ limit.

At a critical separation, the volumes are equal and the stable AdS-Euler solution becomes unstable. If we are careful not to perturb the system significantly and by moving the two Wilson surfaces apart adiabatically, then we may be able to keep the AdS-Euler soluton past the critical value where the AdS-Goldschmidt solution has a smaller volume,
\bea
V_G < V_E
\eea
The partition function in this regime is on the form
\bea
Z &=& D_G e^{-\frac{2N}{\pi} V_G} \(1 + \frac{D_E}{D_G}e^{-\frac{2N}{\pi} \(V_E - V_G\)}\)
\eea
and thus dominated entirely by the AdS-Goldschmidt solution in the large $N$ limit, where it would be basically impossible (that is, probability zero) to preserve the AdS-Euler solution past the critical separation no matter how gently we would move the two spheres away from each other. Quantum fluctuations would destabilize the surface and it would decay to the AdS-Goldschmidt solution through a first-order phase transition right at the critical separation.

\section{Two spherical Wilson surfaces in AdS}
While the full action for the supermembrane in $AdS_7 \times S^4$ is rather complicated, what we need from this action that will be relevant for the classical on-shell value of this action on the saddles, is the bosonic Nambu-Goto action, which is given by
\bea
S &=& \int_0^{\pi} d\theta \int_0^{2\pi} d\varphi \int d\tau \sqrt{g}
\eea
Other terms in the supermembrane action will be important for the quantum fluctuations around the saddle points, which will not be our concern here. In the above membrane action, $g$ denotes the induced membrane metric, which is induced from the AdS$_7$ metric in the Poincare patch
\bea
ds^2 &=& \frac{1}{z^2} \(dz^2 + dt^2 + dx^2 + dy^2 + dr^2 + r^2 \(d\theta^2 + \sin^2\theta d\varphi^2\)\)
\eea
The boundary at $z=\eps$ has the induced boundary metric
\bea
ds^2 &=& \frac{1}{\eps^2} \(dt^2 + dx^2 + dy^2 + dr^2 + r^2 \(d\theta^2 + \sin^2\theta d\varphi^2\)\)
\eea
We will make the following ansatz for the membrane geometry
\bea
t &=& 0\cr
r &=& R(\tau)\cr
x &=& X(\tau)\cr
y &=& 0\cr
z &=& Z(\tau)
\eea
while $\theta,\varphi$ are left as parameters on the surface. This means that we assume that the membrane solution will be spherically symmetric with the induced metric 
\bea
ds^2 &=& \frac{1}{Z(\tau)^2} \(\dot{Z}^2 + \dot{X}^2 + \dot{R}^2\) d\tau^2 + R^2 \(d\theta^2 + \sin^2\theta d\varphi^2\)
\eea
and by this spherical ansatz the membrane action reduces to a one-dimensional action
\bea
S &=& 4 \pi \int d\tau \frac{R^2(\tau)}{Z^3(\tau)} \sqrt{\dot{Z}^2 + \dot{X}^2 + \dot{R}^2}
\eea
where the area of the unit sphere is present just as the prefactor of $4\pi$. 
Let us define the Lagrangian as
\bea
L &=& \frac{R^2}{Z^3} \sqrt{\dot{Z}^2 + \dot{X}^2 + \dot{R}^2}
\eea
Then the conjugate momenta are
\bea
p_Z &=& \frac{R^2}{Z^3} \frac{\dot{Z}}{\sqrt{\dot{Z}^2 + \dot{X}^2 + \dot{R}^2}}\cr
p_X &=& \frac{R^2}{Z^3} \frac{\dot{X}}{\sqrt{\dot{Z}^2 + \dot{X}^2 + \dot{R}^2}}\cr
p_R &=& \frac{R^2}{Z^3} \frac{\dot{R}}{\sqrt{\dot{Z}^2 + \dot{X}^2 + \dot{R}^2}}
\eea
These momenta are subject to the primary constraint
\bea
p_Z^2 + p_X^2 + p_R^2 &=& \frac{R^4}{Z^6}
\eea
The canonical Hamiltonian is vanishing, and instead we have the constraint taking the role of a Hamiltonian as
\bea
H &=& \frac{\lambda}{2} \(p_Z^2 + p_X^2 + p_R^2 - \frac{R^4}{Z^6}\)
\eea
where $\lambda$ is a parameter that we shall not consider as a dynamical variable. 

We have a conserved charge, the dilatation charge, which is given by
\bea
D &=& R p_R + Z p_Z + X p_X
\eea
It generates the scale transformations 
\bea
\delta R &=& c R\cr
\delta Z &=& c Z\cr
\delta X &=& c X
\eea
under which the action is invariant so the dilatation charge is conserved,
\bea
\frac{dD}{d\tau} &=& 0\cr
[H,D] &=& 0
\eea

From the Hamiltonian we derive the following Hamilton equations of motion
\bea
\dot{Z} &=& \lambda p_Z\cr
\dot{X} &=& \lambda p_X\cr
\dot{R} &=& \lambda p_R\cr
\dot{p}_R &=& \lambda \frac{2 R^3}{Z^6}\cr
\dot{p}_Z &=& - \lambda \frac{3 R^4 }{Z^7}\cr
\dot{p}_X &=& 0
\eea
In addition to these, we also need to also impose the constraint equation separately. We can trivially integrate the $p_X$-equation of motion and get
\bea
p_X &=& \alpha
\eea
which is a constant. We can then integrate the $X$-equation of motion to get
\bea
X &=& \lambda \alpha \tau + \beta
\eea
where $\beta$ is another constant. We now need to distinguish between two cases, which are when $\alpha \neq 0$ and when $\alpha = 0$ respectively. For the case when $\alpha \neq 0$ the parameter $\tau$ is nothing but the $x$-coordinate up to a constant rescaling and constant shift and we may then simply put
\bea
\tau &=& x
\eea
such that 
\bea
X(x) &=& x
\eea
We will refer to the solution for this case as the AdS-Euler solution. It is a generalization to AdS space of the catenoid solution that connects two separated and concentrical circles in flat euclidean three-dimensional space that was studied by Leonhard Euler in 1744. 

For the case of $\alpha = 0$ we have $X = \beta$, which is a constant, and we may then take the parameter as $\tau = z$. This gives what we will refer to as the AdS-Goldschmidt solution.

\subsection{The AdS-Goldschmidt solution}
For the AdS-Goldschmidt solution where $\alpha=0$, the constraint equation reads
\bea
p_Z^2 + p_R^2 - \frac{R^4}{Z^6} &=& 0
\eea
We may keep the parameter $\tau$ arbitrary, and then we have the Hamilton equations 
\bea
\dot{Z} &=& \lambda p_Z\cr
\dot{R} &=& \lambda p_R\cr
\dot{p}_Z &=& - \frac{3\lambda R^4}{Z^7}\cr
\dot{p}_R &=& \frac{2\lambda R^3}{Z^6}
\eea
and the dilatation charge
\bea
D &=& R p_R + Z p_Z\cr
&=& \frac{1}{\lambda} \(R \dot{R} + Z \dot{Z}\) 
\eea
which we can rewrite as 
\bea
2\lambda D &=& \frac{d}{d\tau} \(R^2 + Z^2\)
\eea
and this can be integrated to give us
\bea
2\lambda D \tau + \gamma &=& R^2 + Z^2\label{dilatation}
\eea
where $\gamma$ is an integration constant. The constraint is 
\bea
\dot{Z}^2 + \dot{R}^2 &=& \frac{\lambda^2 R^4}{Z^6}
\eea
The remaining equations of motion are
\bea
\frac{1}{\lambda} \frac{d}{d\tau} \(\frac{1}{\lambda} \frac{dZ}{d\tau}\) &=& - \frac{3 R^4}{Z^7}\cr
\frac{1}{\lambda} \frac{d}{d\tau} \(\frac{1}{\lambda} \frac{dR}{d\tau}\)  &=& \frac{2 R^3}{Z^6}
\eea
To simplify the analysis of these equations, we will set $\lambda = 1$ as a gauge choice. Then these equations of motion reduce to
\bea
\ddot{Z} &=& - \frac{3 R^4}{Z^7}\cr
\ddot{R} &=& \frac{2 R^3}{Z^6}
\eea
We can determine the value of the dilation charge $D$ by looking at the maximal value of $Z$ where we shall have $dZ/d\tau = 0$ and $R = 0$. Now we recall that 
\bea
D &=& R p_R + Z p_Z
\eea
and that $p_Z = \lambda \frac{dZ}{d\tau} = 0$ at the tip, thus resulting in 
\bea
D &=& 0
\eea
if we assume that $p_R$ is finite at the tip. To see this, we look at the constraint equation
\bea
p_R^2 + p_Z^2 &=& \frac{R^4}{Z^6}
\eea
Thus at the tip, where $R=0$, this collapses into the condition
\bea
p_R &=& 0\cr
p_Z &=& 0
\eea
implying that $D = 0$. We can now present the AdS-Goldschmidt solution. It is given by 
\bea
R^2 + Z^2 &=& R_b^2\cr
X &=& \pm \frac{L}{2}
\eea
where $R_b$ is the radius of each of the two Wilson surfaces at the boundary as measured by the boundary metric. We see that the AdS-Goldschmidt saddle is nothing but the spherical Wilson surface in \cite{Berenstein:1998ij} that we associate to each spherical Wilson surfaces separately, with seemingly no interaction at all between them, but of course there will be a relatively small interaction between these two Wilson surfaces nonetheless, which comes from supergraviton exchanges that will require a different kind of analysis that we will not study here.

We compute the volume of the AdS-Goldschmidt solution in the parametrization where $z$ is taken as the parameter,
\bea
V_{G} &=& 8\pi \int_{\eps}^{R_b} dz \frac{R^2}{z^3} \sqrt{1+\(\frac{dR}{dz}\)^2}
\eea
We thus compute the volume for the solution at $X = - L/2$ and then multiply this volume by a factor of two to account for the equally large volume solution at $X = L/2$. We shall now plug in the solution 
\bea
R(z) &=& \sqrt{R_b^2 - z^2}
\eea 
and we get the integral
\bea
V_{G} &=& 8\pi R_b  \int_{\eps}^{R_b} dz \frac{\sqrt{R_b^2-z^2}}{z^3}
\eea
It is useful to also present the volume integral that we get upon the coordinate change $z = R_b \sin\theta$. This gives us the integral 
\bea
V_{G} &=& 8 \pi \int_{\eps/R_b}^{\pi/2} d\theta \frac{\cos^2\theta}{\sin^3\theta}
\eea
We now change to the coordinate $u = \sin^2 \theta$, which gives us
\bea
V_{G} = 4 \pi \int_{\epsilon^2/R_b^2}^{1} \frac{\sqrt{1-u}}{u^2} \, du
\eea
The indefinite integral can be computed exactly as 
\bea
\int du \frac{\sqrt{1-u}}{u^2} &=& \ln\(\frac{1+\sqrt{1-u}}{\sqrt{u}}\) - \frac{\sqrt{1-u}}{u} 
\eea
We see that when we insert the upper boundary value $u = 1$ we get a vanishing contribution, so that the whole contribution comes from the lower regularized boundary value at $u = \eps^2/R_b^2$, which gives us
\bea
V_G(\epsilon) = \frac{4\pi R_b^2}{\epsilon^2} + 4\pi\ln\left(\frac{\epsilon}{2R_b}\right) - 2\pi + \mathcal{O}(\epsilon^2)
\eea
which is twice (because we have two spherical Wilson surfaces instead of just one) of the result that was found in \cite{Berenstein:1998ij}.

\subsection{The AdS-Euler solution}
For the AdS-Euler solution we take 
\bea
\tau = x
\eea
and the constraint equation becomes
\bea
\dot{R}^2 + \dot{Z}^2 &=& \frac{\lambda^2 R^4}{Z^6} - 1
\eea
We have the dilatation charge
\bea
D &=& R p_R + Z p_Z + X p_X
\eea
We can rewrite this relation in the form
\bea
D \lambda &=& R \dot{R} + Z \dot{Z} + X \dot{X}\cr
&=& \frac{1}{2} \frac{d}{d\tau} \(R^2 + Z^2 + X^2\)
\eea
that we can integrate to get
\bea
R^2 + Z^2 + x^2 &=& 2 D \lambda x + a^2
\eea
where $a^2$ is an integration constant. We determine $D$ by evaluation at both the end points $x = \pm L/2$ where we assume that $R = R_b$ and $Z = 0$. This gives $D = 0$ and $a^2 = R_b^2 + L^2/4$, and hence
\bea
R^2 + Z^2 &=& a^2 - x^2
\eea
We shall now solve the equations
\bea
\dot{Z}^2 + \dot{R}^2 &=& \frac{\lambda^2 R^4}{Z^6} - 1\cr
R^2 + Z^2 &=& a^2 - x^2
\eea
together. We put
\ben
R(x) &=& \sqrt{a^2-x^2} \cos\theta(x)\cr
Z(x) &=& \sqrt{a^2-x^2} \sin\theta(x)\label{Eulersolution}
\een
and then 
\bea
\dot{R}^2 + \dot{Z}^2 &=& \frac{x^2}{a^2 - x^2} + (a^2 - x^2) \dot{\theta}^2
\eea
By using the primary constraint we obtain the following differential equation for the AdS-Euler solution,
\ben
\dot{\theta}^2 &=& \frac{1}{(a^2 - x^2)^2} \left( \frac{\lambda^2 \cos^4\theta}{\sin^6\theta} - a^2 \right)\label{314}
\een
We take the square root and separate the variables as
\ben
\frac{\sin^3\theta \, d\theta}{\sqrt{\lambda^2 \cos^4\theta - a^2 \sin^6\theta}} &=& \frac{dx}{a^2 - x^2}\label{315}
\een
This we shall now integrate to obtain the Euler solution implicitly via an incomplete elliptic integral,
\bea
\int_0^{\theta(x)} \frac{\sin^3\theta \, d\theta}{\sqrt{\lambda^2 \cos^4\theta - a^2 \sin^6\theta}} &=& \int_{-L/2}^x \frac{dx'}{a^2 - x'^2}\cr
&=& \frac{1}{2a} \(\ln\frac{a+x}{a-x} - \ln\frac{a-\frac{L}{2}}{a+\frac{L}{2}}\)
\eea
We now introduce the parameter 
\bea
\xi &=& \frac{a}{\lambda}
\eea
that enables us to rewrite the relation above in the form
\bea
\xi \int_0^{\theta(x)} \frac{\sin^3\theta \, d\theta}{\sqrt{\cos^4\theta - \xi^2 \sin^6\theta}} &=& \frac{1}{2} \(\ln\frac{a+x}{a-x} - \ln\frac{a-\frac{L}{2}}{a+\frac{L}{2}}\)
\eea
This relation gives the solution $\theta(x)$ implicitly as a function of $x$, which in turn then gives $R(x)$ and $Z(x)$ from (\ref{Eulersolution}) as functions of $x$. We will refer to this as the AdS-Euler solution. 
 
Let us introduce the complete elliptic function 
\bea
F\(\xi\) &=& \xi \int_0^{\theta_0} \frac{\sin^3\theta \, d\theta}{\sqrt{\cos^4\theta - \xi^2 \sin^6\theta}} 
\eea
Here $\theta_0$ is defined such that the square root vanishes, that is
\bea
\cos^2 \theta_0 &=& \xi \sin^3 \theta_0
\eea
At the same time, we have that $\theta_0 = \theta(0)$ is the value of $\theta$ where the bulk solution is turning around as we may deduce from eq (\ref{314}), and by a symmetry argument, we expect this to happen at the mid-point $x = 0$ for the case of two boundary spheres of equal radii that are located at $x = \pm L/2$ respectively. 

Given the above definition of $F$, we now have an equation for $\lambda$,
\bea
F\(\frac{a}{\lambda}\) &=& \frac{1}{2} \ln\frac{a+\frac{L}{2}}{a-\frac{L}{2}}
\eea
We now recall the relation
\bea  
a &=& \sqrt{R_{b}^2 + \frac{L^2}{4}}
\eea
that we now use to rewrite the right-hand side as
\bea
\frac{1}{2}\ln\frac{a+\frac{L}{2}}{a-\frac{L}{2}} &=& \frac{1}{2}\ln\frac{\(a + \frac{L}{2}\)^2}{a^2 - \frac{L^2}{4}}\cr
&=& \ln\frac{\sqrt{R_{b}^2 + \frac{L^2}{4}}+\frac{L}{2}}{R_{b}}
\eea
We now have the equation
\bea
e^F &=& \frac{\sqrt{R_{b}^2 + \frac{L^2}{4}}+\frac{L}{2}}{R_{b}}
\eea
that we can invert to get
\bea
L &=& 2 R_b \sinh(F)
\eea
The function $F(\xi)$ has a maximal value at $\xi = \xi_{max}$ and since the sinh function is a monotous function of $F$ this means that the right-hand side attains a maximal value and henceforth $L$ also attains a maximum possible value 
\bea
L_{max} &=& 2 R_b \sinh\(F\(\xi_{max}\)\)
\eea
beyond which the Euler solution does not exist, be it stable or unstable. 

We can determine the value $\lambda$ from 
\bea
\xi = \frac{a}{\lambda} = \frac{1}{\lambda} \sqrt{R_b^2 + \frac{L^2}{4}} 
\eea
by substituting our explicit solution for $L$, 
\bea
\xi &=& \frac{1}{\lambda} \sqrt{R_b^2 + \frac{\(2 R_b \sinh(F(\xi))\)^2}{4}}
\eea
that fixes the parameter $\lambda$ at the value
\bea
\lambda &=& \frac{R_b}{\xi} \cosh\(F\left(\xi\)\)
\eea
The asymptotic behavors of $F(\xi)$ are obtained in the appendix. In the limit $\xi\rightarrow 0$ we have
\bea
F(\xi) &\approx & \frac{\Gamma\left(\frac{1}{4}\right)^2}{4\sqrt{2\pi}} \, \sqrt{\xi} 
\eea
In the limit $\xi\rightarrow \infty$ we have
\bea
F(\xi) &\approx& \frac{\sqrt{\pi} \, \Gamma\left(\frac{2}{3}\right)}{ \Gamma\left(\frac{1}{6}\right)} \frac{1}{\xi^{1/3}}
\eea
The general shape of $F(\xi)$ is such that for all $\xi>0$ different from the maximum $\xi_{max}$, we have that for any given value of $F$ there are two distinct solutions for $\xi$ that give the same value $F$. At $\xi = \xi_{max}$ these two solutions merge together. 

The graph for $F$ as a function of $\xi$ can be obtained numerically as drawn below from a Python code,
\begin{figure}[H]
    \centering
    \includegraphics[width=0.8\textwidth]{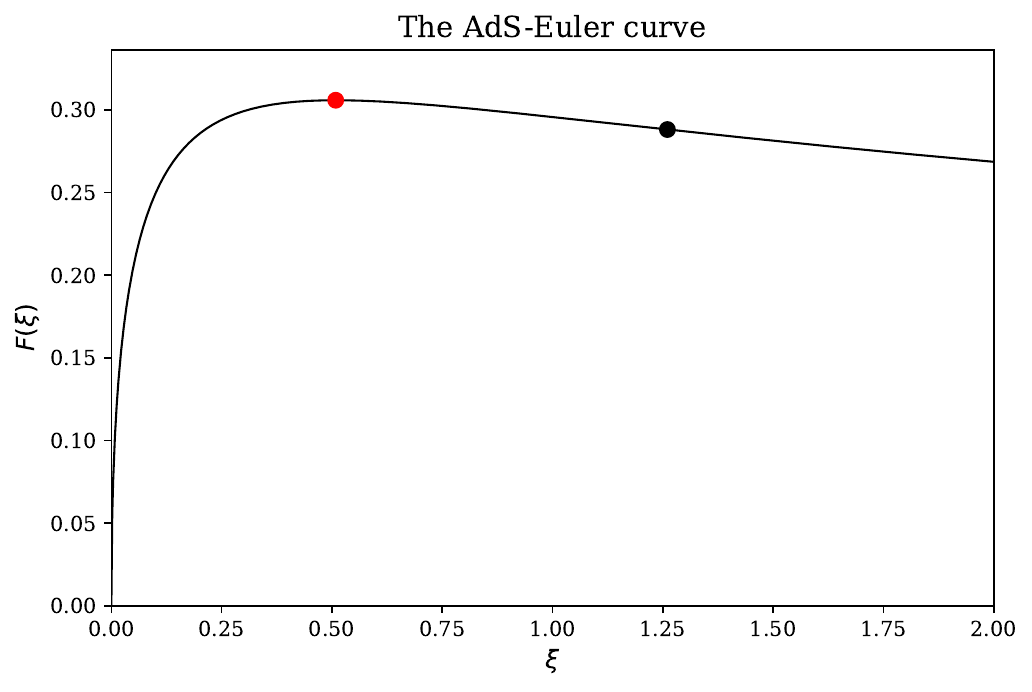}
    \label{fig:sample_plot} 
\end{figure}
Numerically we find that $\xi$ at the maximum of $F$ occurs at 
\bea
\xi_{max} \approx 0.5086
\eea
and that the maximum value there is given by
\bea
F(\xi_{max}) \approx 0.3059
\eea
This point is indicated by a red dot in the graph. The corresponding separation length-to-radius ratio at this maximum value is given by
\bea
\frac{L_{max}}{R_b} = 2 \sinh(F(\xi_{max})) \approx 0.6213
\eea                                                           
For a separation length beyond this separation length, $L > L_{max}$, no AdS-Euler solution exists, neither stable nor unstable ones, which means that the system has to settle down in the AdS-Goldschmidt solution. The black dot in the graph indicates the transition point where the volumes of the AdS-Euler and AdS-Goldschmidt solutions coincide. 

We compute the volume of the AdS-Euler solution in the parametrization where $x$ is taken as the parameter,
\bea
V_E &=& 4 \pi \int_{-L/2}^{L/2} dx \, \frac{R^2}{Z^3} \sqrt{1 + \dot{R}^2 + \dot{Z}^2}
\eea
Substituting the constraint equation $\sqrt{1 + \dot{R}^2 + \dot{Z}^2} = \frac{\lambda R^2}{Z^3}$, the volume becomes
\bea
V_E &=& 4 \pi \lambda \int_{-L/2}^{L/2} dx \, \frac{R^4}{Z^6} = 8 \pi \lambda \int_{-L/2}^{0} dx \, \frac{R^4}{Z^6}
\eea
Using (\ref{Eulersolution}) we get
\bea
\frac{R^4}{Z^6} &=& \frac{1}{a^2-x^2} \frac{\cos^4\theta}{\sin^6\theta}
\eea
and by using (\ref{315}) we obtain the volume as the following integral,
\bea
V_E &=& 8 \pi \int_{\eps/R_b}^{\theta_0} d\theta\frac{\cos^4\theta}{\sin^3\theta \sqrt{\cos^4\theta - \xi^2 \sin^6\theta}} \label{VE}
\eea
where 
\bea
\cos^2 \theta_0 &=& \xi \sin^3 \theta_0
\eea

\section{The difference between Euler and Goldschmidt}
In the limit $\xi \rightarrow 0$ we have $\theta_0 \rightarrow \pi/2$ and we find that $V_E \rightarrow V_G$. This makes it interesting to study the difference between the two volumes, 
\bea
V_E - V_G &=& 8 \pi \int_{\eps/R_b}^{\theta_0} d\theta\frac{\cos^4\theta}{\sin^3\theta \sqrt{\cos^4\theta - \xi^2 \sin^6\theta}} - 8 \pi \int_{\eps/R_b}^{\pi/2} d\theta \frac{\cos^2\theta}{\sin^3\theta}
\eea
Again let us change variable to $u = \sin^2\theta$, 
\bea
V_E - V_G &=& 4\pi \int_{\epsilon^2/R_b^2}^{u_0} du \frac{\sqrt{1-u}}{u^2 \sqrt{1 - \xi^2 \frac{u^3}{(1-u)^2}}} - 4\pi \int_{\epsilon^2/R_b^2}^{1} du \frac{\sqrt{1-u}}{u^2} \cr
&=& 4\pi \int_{\epsilon^2/R_b^2}^{u_0} \frac{\sqrt{1-u}}{u^2} du \left[ \frac{1}{\sqrt{1 - \xi^2 \frac{u^3}{(1-u)^2}}} - 1 \right] - 4\pi \int_{u_0}^{1} du \frac{\sqrt{1-u}}{u^2}
\eea
The second integral has a primitive function that can be expressed in terms of elementary functions,
\bea
\int du \frac{\sqrt{1-u}}{u^2} &=& \log\(1+\sqrt{1-u}\) - \frac{1}{2} \log(u) - \frac{\sqrt{1-u}}{u}
\eea
We then arrive at the following expression for the volume difference
\bea
V_E - V_G &=& I(\xi) + 4\pi \ln\(\frac{1+\sqrt{1-u_0}}{\sqrt{u_0}}\)  - 4\pi \frac{\sqrt{1-u_0}}{u_0} 
\eea
where the first integral can be written in the form
\bea
I(\xi) &=& 4\pi \xi^2 \int_{0}^{u_0} du\frac{u\sqrt{1-u}}{p(u,\xi)+(1-u)\sqrt{p(u,\xi)}}
\eea
where we define
\bea
p(u,\xi) &=& (1-u)^2 - \xi^2 u^3
\eea
While we can not express the integral $I(\xi)$ in terms of elementary functions, we can obtains its asymptotic behaviors for both large and for small values of $\xi$ and we can perform some numerical analysis for intermediate values of $\xi$. 

\subsection{The large-$\xi$ asymptotic behavior}
We will obtain the large $\xi$ asymptotic behavior of the volume difference $V_E- V_G$ by analysing each of the constituent terms participating in the difference formula. First we need to understand the turning point $u_0(\xi)$ that is defined as the smallest positive real root of the polynomial equation
\bea
(1-u_0)^2 - \xi^2 u_0^3 = 0 
\eea
that we can rewrite as
\bea
u_0 &=& \frac{\(1-u_0\)^{2/3}}{\xi^{2/3}}
\eea
where the cubic square root is taken as the positive real root. From this, we see that the asymptotic behavior at infinity for the turning point is 
\bea
u_0(\xi) &=& \frac{1}{\xi^{2/3}} + \O\(\frac{1}{\xi^{4/3}}\)
\eea
This gives us the asymptotic behaviors of the following two terms,
\bea
4\pi \ln\left(\frac{1+\sqrt{1-u_0}}{\sqrt{u_0}}\right) - 4\pi \frac{\sqrt{1-u_0}}{u_0} &=& - 4\pi \xi^{2/3} + \frac{4\pi}{3}\ln(\xi) + 4\pi\ln 2 + 2\pi \cr
&&+ \O\(\frac{1}{\xi^{2/3}}\) 
\eea
This gives a suppression term as the asymptotic behavior $\sim - 4\pi \xi^{2/3}$ at infinity where the log term and the constant terms are completely negligible in comparison.

It now remains to analyze the integral, which upon the substitution $u = \xi^{-2/3}x$ is given by
\bea
4\pi\xi^{2/3} \int_0^1 dx \frac{x \sqrt{1 - \xi^{-2/3} x}}{\(1-\xi^{-2/3}x\)^2 - x^3 + \(1 - \xi^{-2/3} x\) \sqrt{\(1-\xi^{-2/3}x\)^2 - x^3}}
\eea
The leading asymptotic behavior at large $\xi$ is extracted from 
\bea
4\pi\xi^{2/3} \int_0^1 dx \frac{x}{1 - x^3 + \sqrt{1 - x^3}}
\eea
We substistute $t = \sqrt{1-x^3}$ that brings the integral into the form 
\bea
4\pi\xi^{2/3} \frac{2}{3}\int_0^1 dt \(1-t\)^{-1/3} \(1+t\)^{-4/3}
\eea
We now use the following result of the hypergeometric function $\,_2F_1\left(a, 1,c;-1\right)$,
\bea
B(1,c-1) \,_2F_1\left(a, 1,c;-1\right) &=& \int_0^1 dt \(1-t\)^{c-2} \(1+t\)^{-a}
\eea
where $B(a,b)$ is the beta function. By noting that $B(1,2/3) = 3/2$, we now see that 
\bea
4\pi\xi^{2/3} \frac{2}{3} \int_0^1 dt \(1-t\)^{-1/3} \(1+t\)^{-4/3} &=&  4\pi\xi^{2/3} \,_2F_1\left(4/3, 1,5/3;-1\right)
\eea
where
\bea
\,_2F_1\left(4/3, 1,5/3;-1\right) &\approx & 0.5688
\eea
The important point is that this growing term does not overcome the suppression term, but we get a net negative factor given numerically approximatively by  
\bea
\,_2F_1\left(4/3, 1,5/3;-1\right) - 1 &\approx & - 0.4312\label{COEFF}
\eea
We have now found that the large $\xi$ asymptotic behavior of the volume difference is given by
\bea
V_E - V_G &\approx & -  4\pi \cdot 0.4312 \; \xi^{2/3}
\eea
which goes down to negative infinity as $\xi$ goes to plus infinity. This volume difference is a physical quantity, and that it is negative must have a physical interpretation as well. What it signifies is that the AdS-Euler solution gives the far dominant contribution as the absolutely stable solution for large $\xi$.

\subsection{The small-$\xi$ asymptotic behavior}
Again, for small $\xi$, we can obtain the asymptotic behavior for the turning point $u_0 = u_0(\xi)$ by solving the equation
\bea
(1-u_0)^2 &=& \xi^2 u_0^3
\eea
Since for $\xi = 0$ we have $u_0 = 1$, let us expand as
\bea
u_0 &=& 1 - \xi v_0
\eea
for a small $v_0$. The turning point equation in this new variable reads
\bea
v_0^2 &=& (1- \xi v_0)^{3}
\eea
that we solve as
\bea
v_0 &=& 1 - \frac{3}{2} \xi + \O(\xi^2)
\eea
and thus
\bea
u_0 &=& 1 - \xi + \frac{3}{2} \xi^2 + \O(\xi^3)
\eea
We can now expand the terms
\bea
4\pi \ln\left(\frac{1+\sqrt{1-u_0}}{\sqrt{u_0}}\right) - 4\pi \frac{\sqrt{1-u_0}}{u_0}  &=& - \frac{8\pi}{3} \xi^{3/2} + \mathcal{O}\left(\xi^2\right)
\eea
To analyze the small $\xi$ behavior of the integral, we put
\bea
u &=& 1 - \xi x
\eea
which gives $p(u,\xi) = \xi^2 \(x^2 - (1-\xi x)^3\)$ and
\bea
I(\xi) &=& - 4\pi \xi^{3/2} \int_{1/\xi}^{(1-u_0)/\xi} dx \frac{(1-\xi x)\sqrt{x}}{x^2 - (1-\xi x)^3 + x\sqrt{x^2 - (1-\xi x)^3}} 
\eea
The leading asymptotic behavior for small $\xi$ becomes
\bea
I(\xi) &=& 4\pi J \xi^{3/2} + \O\(\xi^2\)
\eea 
where
\bea
J &=& \int_{1}^{\infty} dx \frac{\sqrt{x}}{x^2 - 1 + x\sqrt{x^2 - 1}} 
\eea
Numerically, we find
\bea
J &\approx& 1.5406
\eea
Adding these contributions, we get
\bea
V_E - V_G &=& 4\pi  \(J - \frac{2}{3}\)  \xi^{3/2}\cr
&\approx& 4\pi \cdot 0.8740 \; \xi^{3/2}\label{coeff}
\eea

\section{The Gross-Ooguri phase transition}
Given that the volume difference $V_E - V_G$ as a function of $\xi$ is a continuous function in $[0, \infty]$ and that it is positive for small values of $\xi$ and negative for large values of $\xi$, we conclude that the volume difference will vanish at some critical value $\xi_*$.

We can also numerically obtain the critical value $\xi_*$ to a good approximation. Below we plot the volume difference as a function of $\xi$,
\begin{figure}[H]
    \centering
    \includegraphics[width=0.8\textwidth]{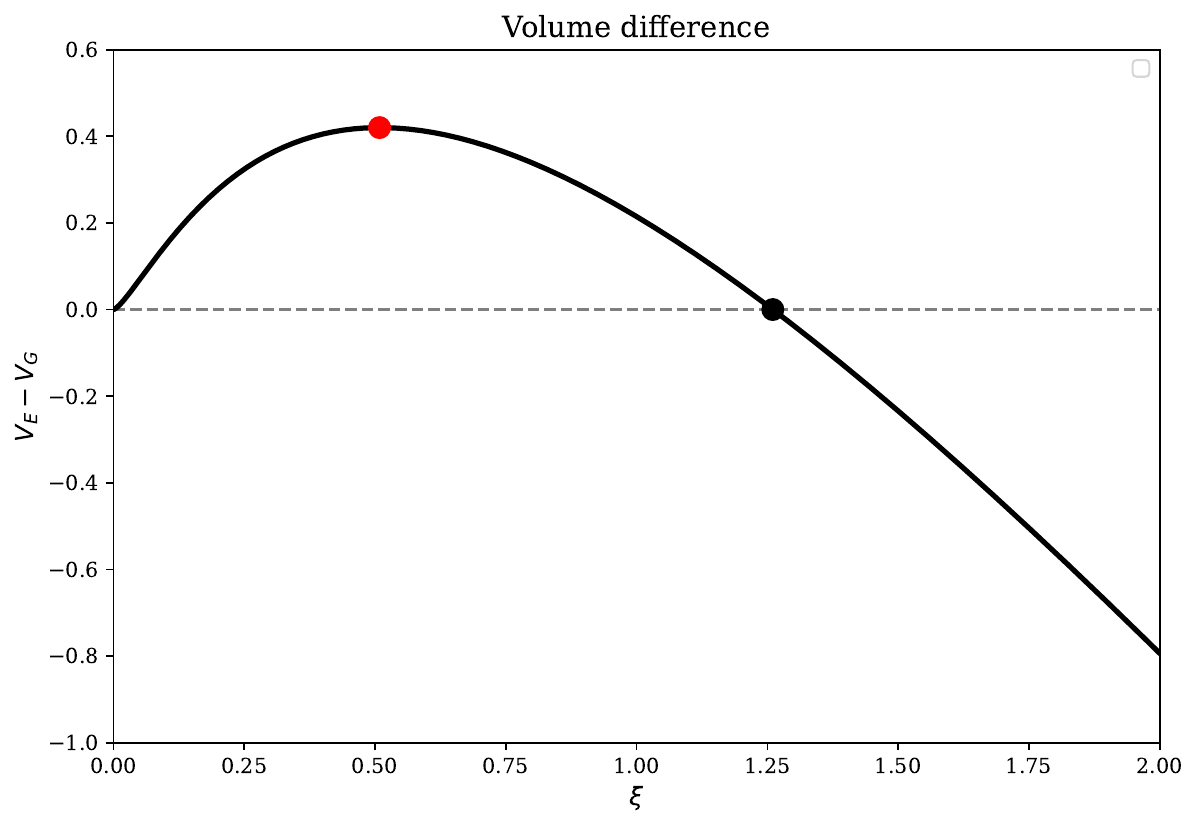}
    \label{fig:sample_plot} 
\end{figure}
The graph shows that the critical Gross-Ooguri phase transition point is at  approximatly
\bea
\xi_* &\approx & 1.2604      
\eea
as indicated by a black dot in the graph where $V_E  = V_G$. The same graph also shows that $V_E - V_G$ reaches its maximum value 
\bea
\(V_E - V_G\)_{max} &\approx  & 0.4201
\eea
at approximatly 
\bea
\xi_{max} &\approx& 0.5086
\eea
We can also plot the volume difference as a function of the separation distance $L$. This graph is plotted below,
\begin{figure}[H]
    \centering
    \includegraphics[width=0.8\textwidth]{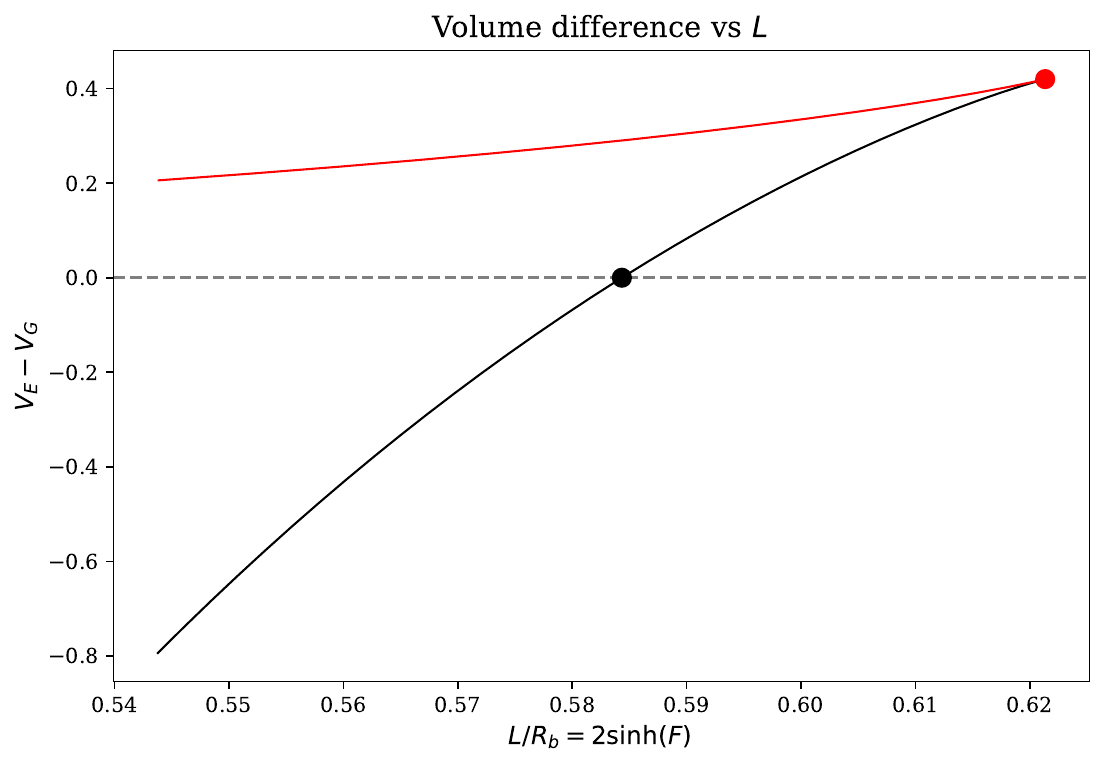}
    \label{fig:sample_plot} 
\end{figure}
The black curve shows how the volume difference goes to minus infinity for small values of $L$. As we increase $L$ we come to the Gross-Ooguri phase transition point at 
\bea
L_* &\approx & 0.5843\; R_b 
\eea
marked with a black dot in the graph. The graph shows that the curve continues beyond this critical value up to a final maximum value 
\bea
L_{max} &\approx & 0.6213 \; R_b
\eea 
marked by a red dot in the graph. Beyond this value, no AdS-Euler solution exists. The graph also shows the unstable AdS-Euler solution that is the red graph for which we have $V_E-V_G>0$ for all values  $0 < L \leq L_{max}$. In the large $N$ limit, where this gravity dual description is valid, the solution will start as a stable AdS-Euler branch as $L$ is increased from zero up to $L_*$. At $L_*$ the solution will change to the AdS-Goldschmidt solution, but the volume will be the same, $V_E = V_G$ at the phase transition point $L_*$, and then as we increase $L>L_*$ we will remain with the AdS-Goldschmidt solution, and as we reach $L_{max}$ nothing dramatic will happen, but we will simply remain with the AdS-Goldschmidt solution all the way up to $L=\infty$. There is only one phase transition point at $L = L_*$ and there we have a first order phase transition reflecting the fact that we have a continous change in the volume. While this is what happens at infinite $N$, for finite values of $N$ the situation gets rather more complicated where we can expect that by including $1/N$ corrections one may be able to see that the AdS-Euler solution may survive to some extent also beyond $L_*$.

\section{At the maximum}
When we look at the graph for $V_E - V_G$ versus $L$ we see a sharp cusp singularity at the maximal value $L_{max}$. This suggests that the maxima of $F$ and $V_E-V_G$ are located at the exact same value $\xi_{max}$. This maximum point is indicated by a single red dot in the graph of $V_E-V_G$ versus $2 \sinh F(\xi)$. 

We can in fact prove that the maxima of $F$ and $V_E -V_G$ occur at the exact same value $\xi_{max}$ by a fully analytic computation. Since $V_G$ is independent of $\xi$ it does not affect the location of the maximum so we can instead of the difference $V_E-V_G$ consider the maximum of $V_E$. We then notice that we have the following decomposition 
\bea
V_E &=& V_{div} + 8 \pi \xi F
\eea
where the divergent part of the volume has been taken out as
\bea
V_{div} &=& 8 \pi \int_{0}^{\theta_0} \frac{\sqrt{\cos^4\theta - \xi^2\sin^6\theta}}{\sin^3\theta} \, d\theta
\eea
The above decomposition follows directly from rewriting the numerator in (\ref{VE}) as $\cos^4 \theta = (\cos^4\theta - \xi^2\sin^6\theta) + \xi^2\sin^6\theta$. A short computation then shows that 
\bea
\frac{d V_{div}}{d\xi} &=& - 8 \pi F
\eea
Taken together, these relations imply that 
\bea
\frac{d V_E}{d\xi} &=& 8 \pi \xi \frac{d F}{d\xi}
\eea
This relation shows that if a maximum value exists at some $\xi_{max}>0$ for $F$ where its derivative vanishes, then the derivative of $V_E$ also has to vanish at $\xi_{max}$. Upon taking the second derivative, we get
\bea
\frac{d^2 V_E}{d\xi^2} &=& 8 \pi \frac{d F}{d\xi} + 8\pi \xi \frac{d^2 F}{d\xi^2}
\eea
This shows that if $F$ attains its maximum value at $\xi_{max}>0$ then the second derivative of the right-hand side is negative, which means that $V_E$ also attains its maximum value at $\xi_{max}$. That $F$ attains a maximum value at some value $\xi_{max}$ would follow solely from the asymptotic behaviors at small and large $\xi$ if we know that $F(\xi)$ is a smooth function. While it seems plausible that $F$ can be a smooth function on the entire half-line $\xi>0$, we do not have a proof of that. 

We have the relation
\ben
\frac{dV_E}{dL} &=& 4\pi \frac{\xi}{a}\label{Mullerhere}
\een
In \cite{Kim:2001td} a similar relation was found in $AdS_5$. We can prove the relation (\ref{Mullerhere}) by the using chain rule
\bea
\frac{dV_E}{d\xi} &=& \frac{dV_E}{dL} \frac{dL}{d\xi} 
\eea
and by noting that
\bea
\frac{dL}{d\xi} &=& 2 a \frac{dF}{d\xi}
\eea
By combining these ingredients, we get
\bea
\frac{dV_E}{d\xi} = \(4\pi \frac{\xi}{a}\) \cdot \(2 a \frac{dF}{d\xi}\) = 8 \pi \xi \frac{dF}{d\xi}
\eea
thus confirming the validity of the relation (\ref{Mullerhere}).

\section{The asymptotics revisisted}
We have obtained asymptotic behaviors for $V_E$ from its integral (\ref{VE}). We can also use the differential relation 
\bea
\frac{d V_E}{d\xi} &=& 8 \pi \xi \frac{d F}{d\xi}
\eea
and from this conclude something about the asymptotic behavior of $V_E$ from the asymptotic behavior of $F$. Not only will this be a nice consistency check, but it will also give us an interesting mathematical identity for the analytic coefficient that governs the large-$\xi$ behavior. 

For large $\xi$ we have the asymptotic behavior
\bea
F(\xi) &=& \frac{C}{\xi^{1/3}}
\eea
where the coefficient is given by 
\bea
C &=& \frac{\sqrt{\pi} \Gamma(2/3)}{\Gamma(1/6)}
\eea
Thus we shall have 
\bea
\frac{d V_E}{d\xi} &=& - \frac{8 \pi C}{3} \xi^{-1/3}
\eea
and accordingly we shall get 
\bea
V_E - V_G &=& 4\pi \cdot C \; \xi^{2/3}
\eea
asymptotically at infinity. By matching the coefficient $C$ here with what we obtained in (\ref{COEFF}) we obtain the following mathematical identity
\bea
\,_2F_1\left(4/3, 1,5/3;-1\right) &=& 1 - \frac{\sqrt{\pi} \Gamma(2/3)}{\Gamma(1/6)}
\eea
This exact analytic value of the hypergeometric function at this particular point seems difficult to derive by direct methods. Let us next turn to the separation length. Asymptotically, for large $\xi$, we find that 
\bea
L &=& 2 R_b \sinh F(\xi)\cr
&\approx& 2 R_b \frac{C}{\xi^{1/3}}
\eea
asymptotically. This shows that we are looking at a very small separation distance as $\xi$ becomes very large. By using this relation, we obtain  
\bea
V_E - V_G &\approx & - \frac{16 \pi R_b^2 C^3}{L^2}
\eea
for small $L$, that is, asymptotically as $L \rightarrow 0$. 

We can repeat the same method for the small $\xi$ asymptotics. We  begin with the asymptotic behavior
\bea
F(\xi) &\approx & c \sqrt{\xi}
\eea
where
\bea
c &=& \frac{\sqrt{\pi}}{4} \frac{\Gamma\(\frac{1}{4}\)}{\Gamma\(\frac{3}{4}\)}
\eea
Then we get
\bea
V_E - V_G &=& \frac{8\pi c}{3} \xi^{3/2}
\eea
and by matching this with (\ref{coeff}) we find the result following exact analytic result,
\bea
J &=& \frac{2}{3} \(\frac{\sqrt{\pi}}{4} \frac{\Gamma\(\frac{1}{4}\)}{\Gamma\(\frac{3}{4}\)} + 1\)
\eea
The separation length behaves as
\bea
L &=& 2 R_b \sinh F(\xi)\cr
&\approx & 2 R_b c \sqrt{\xi}
\eea
in the small $L$ limit, which in turn leads to 
\bea
V_E - V_G &\approx & \frac{\pi}{3 R_b^3 c^2} L^3
\eea
This is the asymptotic behavior of the volume difference for the unstable AdS-Euler solution as $L\rightarrow 0$.

\subsection*{Acknowledgements}
This work was supported in part by Basic Science Research Program through NRF funded by the Ministry of Education (2018R1A6A1A06024977).

\appendix

\section{Asymptotic behaviors of $F(\xi)$}
In this appendix we derive the asymptotic behaviors of 
\bea
F(\xi) &=& \xi \int_0^{\theta_0} d\theta \frac{\sin^3\theta}{\sqrt{\cos^4\theta - \xi^2 \sin^6\theta}}
\eea
for small and for large values of $\xi$ respectively. Here the turning point angle $\theta_0$ satisfies
\ben
\cos^2\theta_0 &=& \xi \sin^3\theta_0 \label{eqe}
\een

\subsection{The asymptotic behavior for $\xi \rightarrow 0$}
We begin with the small $\xi$ behavior of $F(\xi)$ around $\xi = 0$. 
For $\xi = 0$ we have $\theta_0 = \frac{\pi}{2}$. Let us begin by putting  
\bea
\theta_0 &=& \frac{\pi}{2} - u
\eea
so that in this new variable 
\bea
F(\xi) &=& \xi \int_{u_0}^{\frac{\pi}{2}} du \frac{\cos^3 u}{\sqrt{\sin^4 u - \xi^2 \cos^6 u}}
\eea
where 
\bea
\sin^2 u_0 &=& \xi \cos^3 u_0
\eea
For an infinitesimal $u_0$, this reduces to an equation
\bea
(u_0)^2 + \O((u_0)^4) &=& \xi
\eea
that we solve by iteration as
\bea
u_0 &=& \sqrt{\xi} + u^{(1)}_0
\eea
Inserting this back, we get
\bea
\sqrt{\xi} u^{(1)}_0 &=& \O(\xi^2)
\eea
and hence 
\bea
u^{(1)}_0 &=& \O(\xi^{3/2})
\eea
The whole integral now becomes
\bea
F(\xi) &=& \xi \int_{\sqrt{\xi}}^{\frac{\pi}{2}} du \frac{1}{\sqrt{u^4 - \xi^2}} + \O(\xi^{3/2})
\eea
where we have isolated only the leading order behavior coming from the region close to the singularity $u = u_0$, around which we have Taylor expanded the integrand to leading order. Now we substitute $u = v \sqrt{\xi}$ and get
\bea
F(\xi) &=& \sqrt{\xi} \int_1^{\frac{\pi}{2\sqrt{\xi}}} dv \frac{1}{\sqrt{t^4-1}} + \O(\xi^{3/2})
\eea
As we let $\xi$ be infinitesimal the upper limit goes to infinity, and we can approximate the integral with 
\bea
F(\xi) &=& \sqrt{\xi} \int_1^{\infty} dv \frac{1}{\sqrt{v^4-1}} + \O(\xi^{3/2})
\eea
The integral is given by 
\bea
\int_{1}^{\infty} dv \frac{1}{\sqrt{v^4 - 1}} = \frac{\sqrt{\pi}}{4} \frac{\Gamma\(\frac{1}{4}\)}{\Gamma\(\frac{3}{4}\)} 
\eea
We have now found that
\bea
F(\xi) &=& \frac{\sqrt{\pi}}{4} \frac{\Gamma\(\frac{1}{4}\)}{\Gamma\(\frac{3}{4}\)} \sqrt{\xi} + \O(\xi^{3/2})
\eea

\subsection{The asymptotic behavior for $\xi \rightarrow \infty$}
In the limit $\xi \to \infty$ the turning point goes to zero $\theta_0 \to 0$ and accordingly for large $\xi$, by Taylor expanding in a small $\theta_0$ parameter, we see that the turning point is satisfying the equation 
\bea
(\theta_0)^3 \xi &=& 1
\eea
to leading order. The integral to leading order then becomes
\bea
F(\xi) &=& \xi \int_0^{\frac{1}{\xi^{\frac{1}{3}}}} \frac{\theta^3 \, d\theta}{\sqrt{1 - \xi^2 \theta^6}}
\eea
We put $u = \xi^{\frac{1}{3}}\theta$ and the integral becomes
\bea
F(\xi) &=& \frac{1}{\xi^{\frac{1}{3}}} \int_0^1 \frac{u^3 \, du}{\sqrt{1 - u^6}} 
\eea
The integral can be computed by the following steps. We put $u^6 = v$ and get
\bea
F(\xi) &=& \frac{1}{6\xi^{\frac{1}{3}}} \int_0^1 \frac{dv}{v^{\frac{1}{3}} \(1-v\)^{\frac{1}{2}}}
\eea
We now note the definition of the beta function
\bea
B(x,y) = \int_0^1 dv v^{x-1}(1-v)^{y-1} = \frac{\Gamma(x)\Gamma(y)}{\Gamma(x+y)}
\eea
from which we see that 
\bea
F(\xi) = \frac{B\(\tfrac{2}{3}, \tfrac{1}{2}\)}{6\xi^{\frac{1}{3}}} = \frac{\sqrt{\pi}\Gamma\(\tfrac{2}{3}\)}{\Gamma\(\tfrac{1}{6}\)} \frac{1}{\xi^{\frac{1}{3}}}
\eea

\end{document}